\documentclass[pra,twocolumn,groupedaddress,showpacs,floatfix]{revtex4-2}

\usepackage{graphics}
\usepackage{graphicx}
\usepackage{bm}
\usepackage{amsmath,accents}
\usepackage{amsfonts}
\usepackage{amssymb}
\usepackage{latexsym}
\usepackage{color}
\usepackage{bbold}
\usepackage{braket}
\usepackage{float}
\usepackage{hyperref}
\usepackage{extarrows}
\usepackage{dsfont}
\usepackage{subfigure}
\usepackage[rflt]{floatflt}

\def\Tr{\mbox{Tr}}

\begin{document}

\title{Quantum jump metrology in a two-cavity network}
\author{Kawthar Al Rasbi,$^{1,2}$ Almut Beige$^1$ and Lewis A.~Clark$^{3}$}
\affiliation{$^1$The School of Physics and Astronomy, University of Leeds, Leeds LS2 9JT, United Kingdom}
\affiliation{$^2$The Department of Physics, Sultan Qaboos University, Sultanate of Oman}
\affiliation{$^3$Centre for Quantum Optical Technologies, Centre of New Technologies, University of Warsaw, Banacha 2c, 02-097 Warsaw, Poland}

\date{\today}

\begin{abstract}
Quantum metrology enhances measurement precision by utilising the properties of quantum physics. In interferometry, this is typically achieved by evolving highly-entangled quantum states before performing single-shot measurements to reveal information about an unknown parameter.  While this is often the optimum approach, implementation with all but the smallest states is still extremely challenging.  An alternative approach is quantum jump metrology [L. A. Clark {\em et al.}, Phys. Rev. A {\bf 99}, 022102 (2019)] which deduces information by continuously monitoring an open quantum system, while inducing phase-dependent temporal correlations with the help of quantum feedback. Taking this approach here, we analyse measurements of a relative phase in an optical network of two cavities with quantum feedback in the form of laser pulses. It is shown that the proposed approach can exceed the standard quantum limit without the need for complex quantum states while being scalable and more practical than previous related schemes.
\end{abstract}

\maketitle 

\section{Introduction} \label{sec1}

\begin{figure*}[t]
	\includegraphics[width=\textwidth]{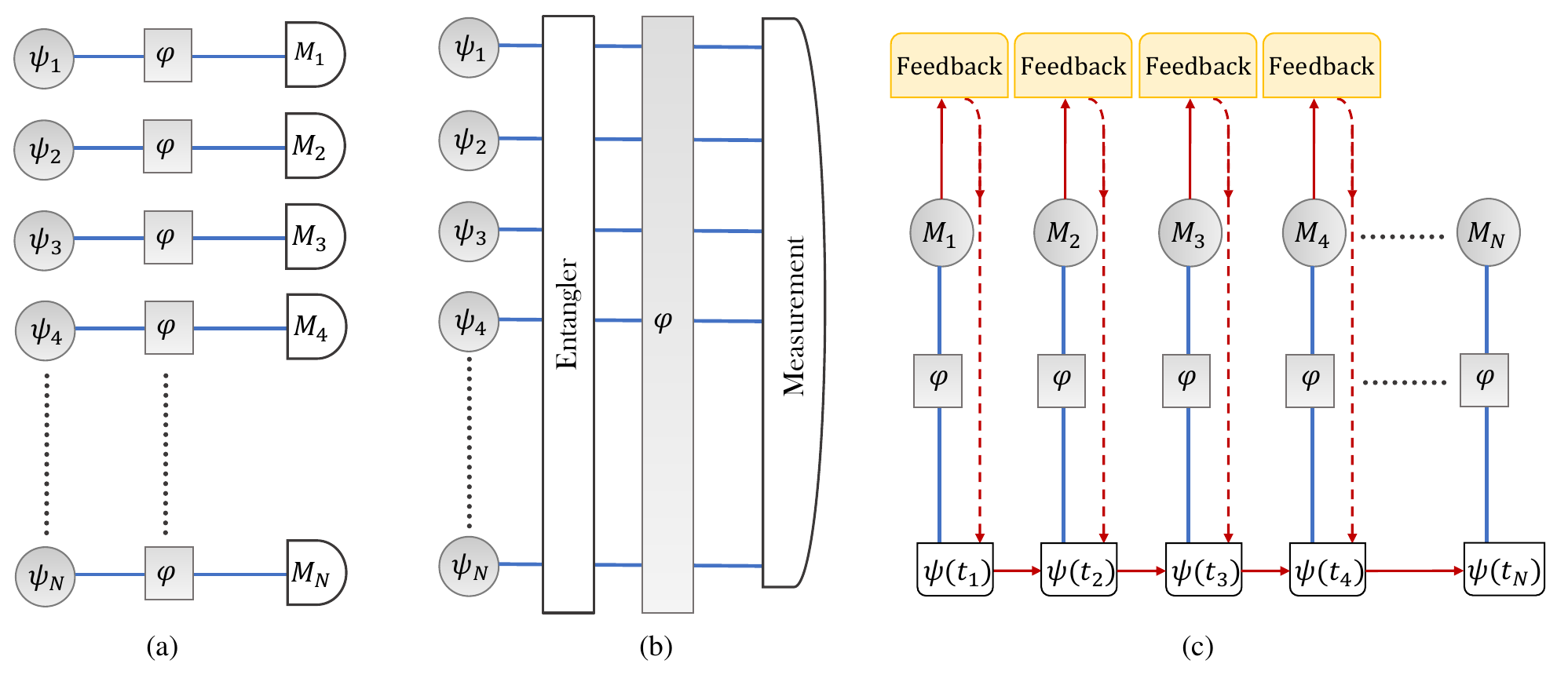}
	\caption{A comparison of three general schemes to achieve an enhancement in estimating an unknown parameter $\varphi$. (a) Classical/uncorrelated scheme consisting of independent and uncorrelated systems where each probe is encoded with the parameter to be determined and then measured.  In such a case the accuracy of measurement is limited by the standard quantum limit. (b) Quantum mechanical scheme utilising entangled states give rise to higher measurement accuracy reaching the Heisenberg limit. (c) Quantum jump metrology scheme where the quantum system is evolved in time inside an instantaneous feedback loop. Sequential measurements on the system allows for an enhancement of the estimation performance beyond the standard quantum limit due to the existence of temporal correlations.}
	\label{fig:scheme}
\end{figure*}

Highly accurate measurements are important for a variety of applications, ranging from probing biological samples \cite{lifetime1,lifetime2,lifetime3} to the detection of gravitational waves \cite{LIGO}. Often, such measurements use light passing through interferometric devices.  Classically, it is well understood how to execute such measurements effectively \cite{kolodynski,yang,Loughridge,scott,ataman,blanco}. One way of increasing their precision is to use higher intensity sources. Sometimes this is not possible, for example, if the object to be probed is fragile or has a short lifetime \cite{lifetime1}. In such cases, precise estimations can only be made after repeating measurements many times. More recently however, it was recognised that one could explore the properties of quantum physics, specifically through quantum metrology \cite{seth,lloyd,Maccone}. These can allow for an increased precision of measurements given the same number of probes.

Suppose an unknown parameter $\varphi$ is determined by performing a measurement with exactly $N$ independent probes, as illustrated in Fig.~\ref{fig:scheme}(a). In this case, the scaling of the uncertainty $\Delta \hat \varphi$ of the estimator with the number of probes $N$ is limited by the standard quantum limit, which tells us that 
\begin{eqnarray} \label{Eq:SQL}
	\left(\Delta \hat{\varphi} \right)^2 & \propto & \frac{1}{N} \, .
\end{eqnarray}
However, for correlated probes, this scaling can be improved.  For maximally correlated probes, the Heisenberg limit may be obtained, which scales with the number of probes as
\begin{eqnarray} \label{Eq:HL}
	\left(\Delta \hat{\varphi} \right)^2 & \propto & \frac{1}{N^2} \, .
\end{eqnarray}
This means that for large $N$, the parameter $\varphi$ can be estimated with the same precision with fewer probes.  Determining how to obtain and utilise such correlated probes has thus been an active area of research.

As illustrated in Fig.~\ref{fig:scheme}(b), one solution is to evolve entangled quantum states in a $\varphi$-dependent fashion followed by a collective measurement of their state \cite{HollandB,CavesC,Bondurant,DowlingNOON,BerryW,Vidrighin,Xiang,Afek,Nagata}. However, a typical problem with this approach is that it is difficult to implement. For example, it has been shown that so-called N00N states, which are highly-entangled $N$-particle states, are optimal for quantum interferometry experiments \cite{DowlingNOON,Afek,Nagata,LeeNoon,Mitchell,PrydeNoon,IsraelNoon}. But reliably obtaining a reasonably large $N$ in the laboratory to realise the above-described enhancement remains extremely challenging \cite{Gilbert,ThomasP}. Using entanglement is not the only way of enhancing measurement precision however \cite{BoixoN,boixo,Napolitano,Atta,Higgins,braun}. Other methods include using effects such as non-linearities. Once again though, non-linearities are often hard to implement experimentally, particularly when processing information with light. 

In this paper we adopt an alternative approach and use {\em quantum jump metrology} \cite{lewisstokes,lewisbeige} to improve the precision of the estimation of an unknown parameter $\varphi$. Quantum jump metrology does not require highly entangled quantum states nor the presence of non-linear optical elements and is therefore relatively easy to implement. Its basic idea is to deduce information about $\varphi$ by monitoring the quantum jump based output statistics of an open quantum system.  To ensure that the dynamics of the individual quantum trajectories of the system depend on $\varphi$, we use quantum feedback \cite{wiseman}, which is triggered by certain measurements outcomes, as illustrated in Fig.~\ref{fig:scheme}(c). Using the dynamics of open quantum systems, in particular in continuously monitored systems, to infer information about an unknown parameter recently received a lot of attention in the literature \cite{Shabani,Beau,Albarelli,Haase}.

At this point it is useful to note the generality of the form of the $N$ `probes' in Eqs.~(\ref{Eq:SQL}) and (\ref{Eq:HL}).  Typically, these can be imagined as individual particles, or some dimensionality of the system.  However, the value $N$ can also be interpreted as the query complexity of the system, i.e.~the number of incompressible steps \cite{Zwierz,lewisstokes}. Hence, while in a single-shot setup, the number of particles is clearly the relevant resource $N$, for a continuously monitored system, $N$ may instead relate to the number of times the system is `probed'.  Considering a measurement scheme for which the overall number of time steps constitutes the relevant resource allows us to exploit non-classical time correlations instead of entanglement to go beyond the standard quantum limit. The method of using non-classical time correlations in quantum technology applications currently attracts a lot of attention in the literature \cite{HQMM,HQMM2,Boots,Tian}. As we shall see below, the measurement observable that we consider in this paper is the relative number of trajectories with multiple successive emission which generate an above-threshold photon number in a given time interval.

\begin{figure*}[t]
	\includegraphics[width=\textwidth]{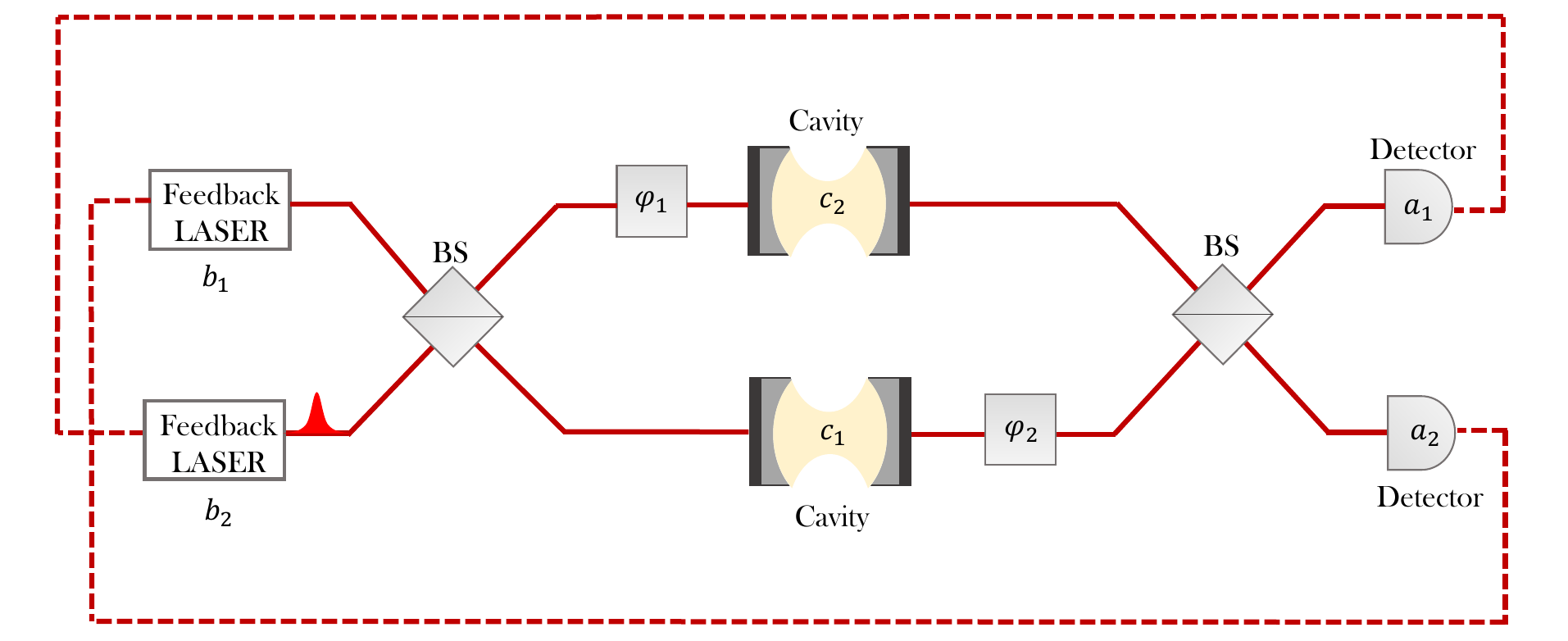}
	\caption{Two optical cavities are monitored through a linear optics network with photon detectors.  Upon the detection of a photon, quantum feedback is triggered and applied to the cavities, also through a linear optics network.   Throughout this paper, we assume the feedback acts instantaneously after a photon detection. In this diagram and our subsequent analysis of the system, we consider the specific case of the dynamics presented in Sec.~\ref{sec2}, where feedback is triggered only in mode $b_2$ from a photon detection in detector 1 and only in mode $b_1$ from a photon detection in detector 2.  The aim of the scheme is to measure the phase difference $\varphi = \varphi_1 - \varphi_2$ from the photon statistics in the detectors.}
	\label{fig:setup}
\end{figure*}

Utilising temporal correlations, Ref.~\cite{lewisstokes} introduced a quantum metrology scheme for measuring the phase difference between a coherent state prepared inside an optical cavity and a laser providing feedback pulses. In this paper, we propose a closely related scheme, which is much more practical and easier to implement, and instead measures the difference $\varphi $ between two phases $\varphi_1$ and $\varphi_2$ corresponding to two different pathways through a linear optics setup, as shown in Fig.~\ref{fig:setup}. Nevertheless, the scheme that we propose here too relies only on coherent states, yet it is capable of producing correlated photon statistics and thus surpassing the standard scaling due to the presence of quantum feedback. Hence, as well as demonstrating a simple scheme with enhanced sensing capabilities, our proposal also demonstrates the power of using quantum feedback to induce quantum effects, even in `classical-like' states such as coherent states.

The use of quantum feedback has found a variety of applications not only in quantum metrology \cite{lewisstokes, lewisbeige}, but also in quantum error correction and noise reduction \cite{qerror}, quantum state stabilisation \cite{sstab}, entanglement control \cite{fentg} and in implementing Hidden Quantum Markov Models \cite{HQMM,HQMM2}.  Moreover, it has recently been shown that quantum feedback can lead to ergodicity breaking in quantum optical systems \cite{lewismaybee}. This can again be achieved even when using only coherent states and feedback in the form of displacements of the field, thus only requiring relatively simple technology to implement. 

Intuitively, it is easy to see how quantum feedback can lead to time correlations in the bath statistics of an open quantum system.  Consider a quantum optical system that emits a photon at a time $t_1$.  Then as the system is perturbed by the feedback the emission probability for another photon is altered.  Thus, the emission at time $t_2$ is correlated with the emission at $t_1$. If this feedback depends on the unknown parameter $\varphi$, these correlations can be used to gain information for its estimation. Hence it is not surprising that quantum feedback is a powerful tool for quantum sensing applications. 

This paper is structured as follows. In Section~\ref{sec2}, we introduce the notation and the basic theoretical tools for the modelling of the linear optics cavity network shown in Fig.~\ref{fig:setup}. Afterwards, in Section~\ref{sec3}, we introduce the measurement scheme that we propose in this paper to sense an unknown phase shift $\varphi$. In Section~\ref{sec4}, we analyse the capabilities of our quantum jump metrology scheme and present numerical results.  Finally, we summarise our findings in Section~\ref{sec5}.

\section{A two-mode cavity network with quantum feedback} \label{sec2}

In this section, we review the main tools for the theoretical modelling of the experimental setup shown in Fig.~\ref{fig:setup}. As we simply consider cavities subject only to laser driving in the form of pulses, resulting in displacements of the cavity field, the system remains always in a coherent state. In the following we have a closer look at the dynamics of these coherent states under the condition of no photon emission and in case of an emission. In addition, we introduce quantum optical master equations that can be used for the prediction of ensemble expectation values.

\subsection{Multi-mode Coherent States and Transformation Matrices} \label{notation}

As mentioned already above, in this paper we consider a network of two leaky optical cavities that are always kept in a coherent state. In Fock state representation, the coherent state $\ket{\gamma_i}$ of cavity $i$ can be solely parametrised by a complex number $\gamma_i$ such that 
\begin{eqnarray}
	\ket{\gamma_i} & = & \exp \left(-\frac{|\gamma_i|^2}{2} \right) \sum\limits_{n_i=0}^\infty \frac{\gamma_i^{n_i}}{\sqrt{n_i\!}} \ket{n_i} \, .
\end{eqnarray}
Here $|n_i \rangle$ is the Fock state with exactly $n_i$ photons in cavity $i$. If $c_i$ is the annihilation operator for a single photon in cavity $i$ with $c_i \ket{n_i} = \sqrt{n_i} \ket{n_i-1}$, then one can show that $c_i \ket{\gamma_i} = \gamma_i \ket{\gamma_i}$. Using this notation, the state $\ket{\psi (t)} $ of both cavities at time $t$ is always of the form
\begin{eqnarray}
\ket{\psi (t)} &=& \bigotimes_{i=1,2} \ket{\gamma_i (t)} 
\end{eqnarray}
and its dynamics can be modelled simply by tracking two complex numbers $\gamma_i(t)$. Hence we can also express the state of the two cavities as
\begin{eqnarray}
	{\underline \gamma} (t) &=& \left( \begin{array}{c} \gamma_1 (t) \\ \gamma_2 (t) \end{array} \right) \, .
\end{eqnarray}
In the following, we adopt a vector and matrix notation for convenience when considering photon counting and quantum feedback processes. 

The experimental setup in Fig.~\ref{fig:setup} contains phase shifters and beamsplitters.  Hence quantum feedback pulses do not perturb the cavities directly. Similarly, photons arriving at a detector do not come directly from only a single cavity. To take this into account more easily, we denote the annihilation operator of the field mode seen by detector $i$ in the following by $a_i$ and the annihilation operator of the field mode affected by laser $i$ by $b_i$. With respect to these alternative modes, the state $|\psi(t) \rangle$ of the cavities is given by the complex vectors 
\begin{eqnarray}
	{\underline \alpha(t)}  =  \left( \begin{array}{c} \alpha_1(t) \\ \alpha_2(t) \end{array} \right) \, , ~~	
	{\underline \beta(t)}  =  \left( \begin{array}{c} \beta_1(t) \\ \beta_2(t) \end{array} \right) 
\end{eqnarray}
with the complex numbers $\alpha_i(t)$ and $\beta_i(t)$ such that $a_i |\alpha_i (t) \rangle = \alpha_i(t) \, |\alpha_i(t) \rangle$ and $b_i |\beta_i(t) \rangle = \beta_i(t) \, |\beta_i(t) \rangle$. To switch from one representation of the cavity network in Fig.~\ref{fig:setup} to another, we define transformation matrices $M_{yx}$ with 
\begin{eqnarray}
	\left( \begin{array}{c} y_1 \\ y_2 \end{array} \right) &=& M_{yx} \left( \begin{array}{c} x_1 \\ x_2 \end{array} \right) \, ,
\end{eqnarray}
where $x_i,y_i = \alpha_i(t) , \beta_i (t), \gamma_i(t) $ for the coefficients of the vectors and where $x,y = a,b,c$ for the subscripts of the transition matrices $M_{yx}$. Below we have a closer look at these matrices, which describe the effect of the beamsplitters and the phase shifters shown in Fig.~\ref{fig:setup}. If we define the transformation matrices
\begin{eqnarray}
	&& S_{\rm BS} =\frac{1}{\sqrt 2} \begin{pmatrix} 1 & {\rm i} \\ {\rm i} & 1 \end{pmatrix} \, , ~~
	S_{\varphi_1} = \begin{pmatrix} 1 & 0 \\ 0 &{\rm e}^{{\rm i}\varphi_1} \end{pmatrix} \, , \nonumber \\
	&& S_{\varphi_2} = \begin{pmatrix} {\rm e}^{{\rm i}\varphi_2}  & 0 \\ 0 & 1  \end{pmatrix} \, , 
\end{eqnarray}
they take the form 
\begin{eqnarray}
	M_{cb} &=& S_{\varphi_1} S_{\rm BS} =  \frac{1}{\sqrt2} \begin{pmatrix} 1& {\rm i} \\ {\rm i} \, {\rm e}^{{\rm i}\varphi_1} & {\rm e}^{{\rm i}\varphi_1} \end{pmatrix} \, , \nonumber \\
	M_{ac} &=& S_{\rm BS} S_{\varphi_2} = \frac{1}{\sqrt2} \begin{pmatrix}   {\rm e}^{{\rm i}\varphi_2} & {\rm i} \\ {\rm i} \, {\rm e}^{{\rm i}\varphi_2} &1 \end{pmatrix} \, . \nonumber \\
	M_{ab} &=& M_{ac} M_{cb} =  \frac{1}{2} \begin{pmatrix} - {\rm e}^{{\rm i}\varphi_1} + {\rm e}^{{\rm i}\varphi_2} & {\rm i} \left(  {\rm e}^{{\rm i}\varphi_1} + {\rm e}^{{\rm i}\varphi_2} \right) \\ {\rm i} \left(  {\rm e}^{{\rm i}\varphi_1} + {\rm e}^{{\rm i}\varphi_2} \right) & {\rm e}^{{\rm i}\varphi_1} - {\rm e}^{{\rm i}\varphi_2} \end{pmatrix} \, . \nonumber \\
\end{eqnarray}
These matrices can now be used to model the dynamics of the system in the different bases. 

\subsection{The effect of quantum feedback} \label{ABC}

Suppose an instantaneous strong laser pulse is applied directly to cavity $i$. Then the effect of this operation on the coherent state $|\gamma_i(t) \rangle$ of mode $i$ in the $c$ basis can be described by a displacement operator of the form
\begin{eqnarray}
	D_i^{(c)}(\beta) & = & \exp \left(\beta \, c_i^\dagger - \beta^* \, c_i \right) 
\end{eqnarray}
where $\beta$ is a complex number that describes the strength and phase of the feedback pulse and can assume any value. Taking this into account, one can show that the result is a change such that 
\begin{eqnarray}
	\gamma_i(t) & \rightarrow & \gamma_i (t) + \beta \, .
\end{eqnarray}
However, in the experimental setup in Fig.~\ref{fig:setup}, quantum feedback does not trigger a laser pulse that disturbs the cavities directly. Instead, because of the presence of a beamsplitter, each laser pulse usually affects the field in both cavities. 

For simplicity, we take the feedback strengths of the laser pulses as constant in time, although it could be made time dependent for further generality. This allows us to model the effect of the feedback by four complex numbers $\beta_i^{(d)}$ which characterise the quantum feedback strength generated by laser $i$ upon detection of a photon in detector $d$ with $d=1,2$. For convenience we arrange these numbers into two vectors
\begin{eqnarray}
	{\underline \beta}^{(d)} & = & \left( \begin{array}{c} \beta_1^{(d)} \\ \beta_2^{(d)} \end{array} \right) \, . 
\end{eqnarray}
Given that the feedback is triggered by the detection of a photon in detector $d$, we observe the following effect of quantum feedback on the state $\underline{\gamma}(t)$ of the cavities, 
\begin{eqnarray}
	\underline{\gamma}(t) & \rightarrow & \underline{\gamma} (t) + M_{cb} \, \underline{\beta}^{(d)}  \, .
\end{eqnarray}
Alternatively, in the basis of the detector modes, the state of the cavities changes such that 
\begin{eqnarray}
	\underline{\alpha} (t) &\rightarrow& \underline{\alpha} (t) + M_{ab} \, \underline{\beta}^{(d)} 
\end{eqnarray}
These equations provide a complete description of the quantum feedback.

\subsection{Master equations and Quantum Jump Approach}

Next we study the effect of the possible leakage of photons through the cavity mirrors on the state of the resonator fields. Because of the presence of spontaneous photon emission, the calculation of expectation values for ensemble averages requires the introduction of a density matrix $\rho$.  In what follows, we work in the detector basis $a_d$ and define all subsequent evolutions and probabilities in terms of this, as it is most convenient for the numerical implementations that follow, although these quantities could in principle be calculated in any basis.  For example, in the absence of quantum feedback, $\rho$ evolves such that 
\begin{eqnarray}
	\dot{\rho} & = & \sum\limits_{d=1,2}  \kappa_d \, a_d \, \rho \, a_d^\dagger - \frac{1}{2}  \kappa_d \, \left[ a^\dagger_d a_d , \rho \right]_+  \, ,
\end{eqnarray}
where $\kappa_d$ denotes the spontaneous decay rate of a single photon in the $a_d$ mode and $\left[ \cdot , \cdot \right]_+$ denotes the anti-commutator.  In the presence of quantum feedback, this master equation changes into
\begin{widetext}
\begin{eqnarray} \label{master}
	\dot{\rho} & = & \sum\limits_{d=1,2}  \kappa_d \, D^{(a)}_2 (\beta^{(d)}_2) D^{(a)}_1 (\beta^{(d)}_1) a_d \, \rho \, a_d^\dagger D^{(a) \dagger}_1 (\beta^{(d)}_1) D^{(a) \dagger}_2 (\beta^{(d)}_2) - \frac{1}{2}  \kappa_d \, \left[ a^\dagger_d a_d , \rho \right]_+ \, .
\end{eqnarray}
\end{widetext}
This equation takes into account that quantum feedback can be interpreted as a modification of the system-bath coupling, thereby resulting in a transformation of the Lindblad operators.  The reason for this change of operators is that the emission of a photon is immediately followed by the application of the feedback pulse(s) \cite{lewisstokes,lewismaybee}.  In obtaining this equation, we have made the standard quantum optical approximations of Markovianity and a rotating wave approximation, while assuming classical driving fields for the laser pulses.  More detailed discussions of  the dynamics of density matrices in the presence of quantum feedback can be found for example in Refs.~\cite{lewisstokes,lewisbeige,wiseman,lewismaybee}. In the following, we have a closer look at an unravelling of the above ensemble dynamics into individual quantum trajectories. These can be studied analytically relatively easily, especially if the cavities are initially prepared in a pair coherent state, like $\underline{\gamma} (0)$. 

\subsubsection{The no-photon time evolution}

To obtain the conditional no-photon evolution, we write the master equation in Eq.~(\ref{master}) as
\begin{widetext}
\begin{eqnarray} \label{18}
	\dot{\rho} & = &  -\frac{\rm i}{\hbar} \left(H_{\rm cond} \rho - \rho H_{\rm cond}^\dagger \right) + \sum\limits_{d=1,2} \kappa_d \,D^{(a)}_2 (\beta^{(d)}_2) D^{(a)}_1 (\beta^{(d)}_1) a_d \, \rho \, a_d^\dagger D^{(a) \dagger}_1 (\beta^{(d)}_1) D^{(a) \dagger}_2 (\beta^{(d)}_2) \, .
\end{eqnarray}
\end{widetext}
with the conditional Hamiltonian $H_{\rm cond}$ given by
\begin{eqnarray}
	H_{\rm cond} &=& -\frac{\rm i}{2} \hbar \sum\limits_{d=1,2} \kappa_d \, a_d^\dagger a_d \, .
\end{eqnarray}
While the last terms in Eq.~(\ref{18}) describe dynamics of the subensembles of systems with a photon detection in output port $d$, the first two terms describe the subensemble without an emission. In other words, the non-Hermitian Hamiltonian $H_{\rm cond}$ is the generator for time evolution of the experimental setup in Fig.~\ref{fig:setup} conditioned on no photon emission. The corresponding time evolution operator 
\begin{eqnarray}
U_{\rm cond}(t,t_0) &=& \exp \left(-\frac{{\rm i}}{\hbar} H_{\rm cond} (t - t_0) \right)
\end{eqnarray}
reduces the norm of state vectors and can be used to calculate the probability $P_{00}(t,t_0)$ for no photon detection in both detectors in a time interval $[t_0,t]$.  This probability equals
\begin{eqnarray} \label{P0}
P_{00}(\Delta t) &=& \| U_{\rm cond}(t,t_0) |\psi(t_0) \rangle \|^2
\end{eqnarray}
for a given initial state $ |\psi(t_0) \rangle$ and $\Delta t = t - t_0$. For example, given an initial pair coherent state $ |\psi(t_0) \rangle = \ket{\alpha_1}\ket{\alpha_2}$ with respect to the modes $a_1$ and $a_2$ seen by the detector, one can show that  
\begin{eqnarray}
	U_{\rm cond} (t,t_0) \ket{\alpha_1}\ket{\alpha_2} & = & \exp\left(\frac{-|\alpha_1|^2}{2} \left(1 - {\rm e}^{-\kappa_1 \Delta t} \right)\right) \nonumber \\
	&& \times \exp\left(\frac{-|\alpha_2|^2}{2} \left(1 - {\rm e}^{-\kappa_2 \Delta t} \right)\right) ~~~~ \nonumber \\
	&& \times \ket{\alpha_1 {\rm e}^{-\frac{1}{2} \kappa_1 \Delta t}} \ket{\alpha_2 {\rm e}^{-\frac{1}{2} \kappa_2 \Delta t}} \, .
\end{eqnarray}
Hence, using the notation introduced in Section~\ref{notation}, we can summarise the effect of the no-photon time evolution of the field inside the cavities as $\underline{\alpha}(t) = M_{00} (t) \underline{\alpha}(t_0)$ with
\begin{eqnarray} \label{M0}
	M_{00} (\Delta t) &=& \begin{pmatrix} {\rm e}^{-\frac{1}{2}\kappa_1 \Delta t }&0 \\ 0& {\rm e}^{-\frac{1}{2}\kappa_2 \Delta t} \end{pmatrix} \, .
\end{eqnarray}
The probability of such an evolution occurring is
\begin{eqnarray} \label{pss3}
	P_{00}(\Delta t) &= & \exp \left[ - |\alpha_1(t_0)|^2 \left(1-{\rm e}^{-\kappa_1 \Delta t} \right) \right] \nonumber \\
	&& \times \exp \left[ - |\alpha_2(t_0)|^2 \left(1-{\rm e}^{-\kappa_2 \Delta t} \right) \right] \, ,
\end{eqnarray}
due to Eq.~(\ref{P0}).

\subsubsection{Photon emission probabilities}

Next we calculate the probabilities of photon emission in a time interval of length $\Delta t$. Having a closer look at the two factors in Eq.~(\ref{pss3}), we see that the probability for an individual detector mode $i$ not to detect a photon equals
\begin{eqnarray} \label{Eq:P0}
	P_{0}^{(i)}(\Delta t) & = & \exp \left[ - |\alpha_i(t_0)|^2 \left(1-{\rm e}^{-\kappa_i \Delta t} \right) \right] \, .
\end{eqnarray}
Moreover, we know that the probability to find at least one photon in detector $i$ is given by $1-P^{(i)}_0(\Delta t)$. Thus the probability for no photon in detector 1 and at least one photon in detector 2 equals 
\begin{eqnarray} \label{Eq:P01}
	P_{01} (\Delta t) & = & \exp \left[ - |\alpha_1(t_0)|^2 \left(1-{\rm e}^{-\kappa_1 \Delta t} \right) \right] \nonumber \\
	&& \times \left(1 - \exp \left[ - |\alpha_2(t_0)|^2 \left(1-{\rm e}^{-\kappa_2 \Delta t} \right) \right] \right) \, . \nonumber \\
\end{eqnarray}
Analogously,
\begin{eqnarray} \label{Eq:P10}
	P_{10} (\Delta t) & = & \left(1 - \exp \left[ - |\alpha_1(t_0)|^2 \left(1-{\rm e}^{-\kappa_1 \Delta t} \right) \right] \right) \nonumber \\
	&& \times \exp \left[ - |\alpha_2(t_0)|^2 \left(1-{\rm e}^{-\kappa_2 \Delta t} \right) \right] 
\end{eqnarray}
is the probability for at least one photon in detector 1 and no photon in detector 2. To cover all possibilities (i.e.~to have probabilities that sum to unity), we consider in the following also the probability 
\begin{eqnarray} \label{Eq:P11}
	P_{11} (\Delta t) & = & \left(1 - \exp \left[ - |\alpha_1(t_0)|^2 \left(1-{\rm e}^{-\kappa_1 \Delta t} \right) \right] \right) \nonumber \\
	&& \times \left(1 - \exp \left[ - |\alpha_2(t_0)|^2 \left(1-{\rm e}^{-\kappa_2 \Delta t} \right) \right] \right) \nonumber \\
\end{eqnarray}
for the case in which at least one photon has been emitted into both detector modes. 

For relatively short time intervals $\Delta t$, the presence of two photons in one detector becomes negligible and the probabilities $P_{01}(\Delta t)$, $P_{10}(\Delta t)$ and $P_{11} (\Delta t)$ become the probabilities to have exactly one photon in mode 2, exactly one photon in mode 1 and exactly one photon in each mode, respectively. This applies to a good approximation, so long as $\Delta t \ll \kappa |\alpha_{i}|^2$.  In fact, the probability $P_{11}(\Delta t)$ would also be negligible by the same argument, as its first non-zero term when expanded is ${\cal O}(\Delta t^2)$, compared to ${\cal O}(\Delta t)$ for single photon emissions. We nevertheless persist in keeping this term to maintain probabilities summing to exactly one. Now the changes of the state vector ${\underline \alpha}(0)$ of the cavity fields can be described by transformation operators $M_{ij} (\Delta t)$ such that 
$\underline{\alpha}(t) = M_{ij} (\Delta t) \, \underline{\alpha}(0)$. With the above approximation in mind, the $M_{ij} (\Delta t)$ are given by
\begin{eqnarray}
	M_{01}(\Delta t ) \, \underline{\alpha}(0) & = & M_{00}(\Delta t ) \left(\underline{\alpha}(0) + \underline{\beta}^{(2)} \right) \nonumber \\
	M_{10}(\Delta t ) \, \underline{\alpha}(0) & = & M_{00}(\Delta t ) \left(\underline{\alpha}(0) + \underline{\beta}^{(1)} \right) \nonumber \\
	M_{11}(\Delta t ) \, \underline{\alpha}(0) & = & M_{00}(\Delta t ) \left(\underline{\alpha}(0) + \underline{\beta}^{(1)} + \underline{\beta}^{(2)} \right)  ~~~
\end{eqnarray}
with $ M_{00}(\Delta t )$ given in Eq.~(\ref{M0}).  Specifically, we assume that the cavity field freely decays in the time interval of size $\Delta t$, but is first displaced by the feedback. So long as $\Delta t$ is sufficiently small, i.e.~as long as $\Delta t \ll \kappa |\alpha_i|^2$ as specified above, the exact moment of the feedback pulse does not significantly change the evolution of the system. We now have a complete toolbox for modelling quantum trajectories through piecewise evolution of the system, as suggested by standard quantum jump methods \cite{Reset,Molmer,Carmichael}.

\section{General dynamics and temporal correlations} \label{sec3}

\begin{figure}
	\centering
	\includegraphics[width=0.5 \textwidth]{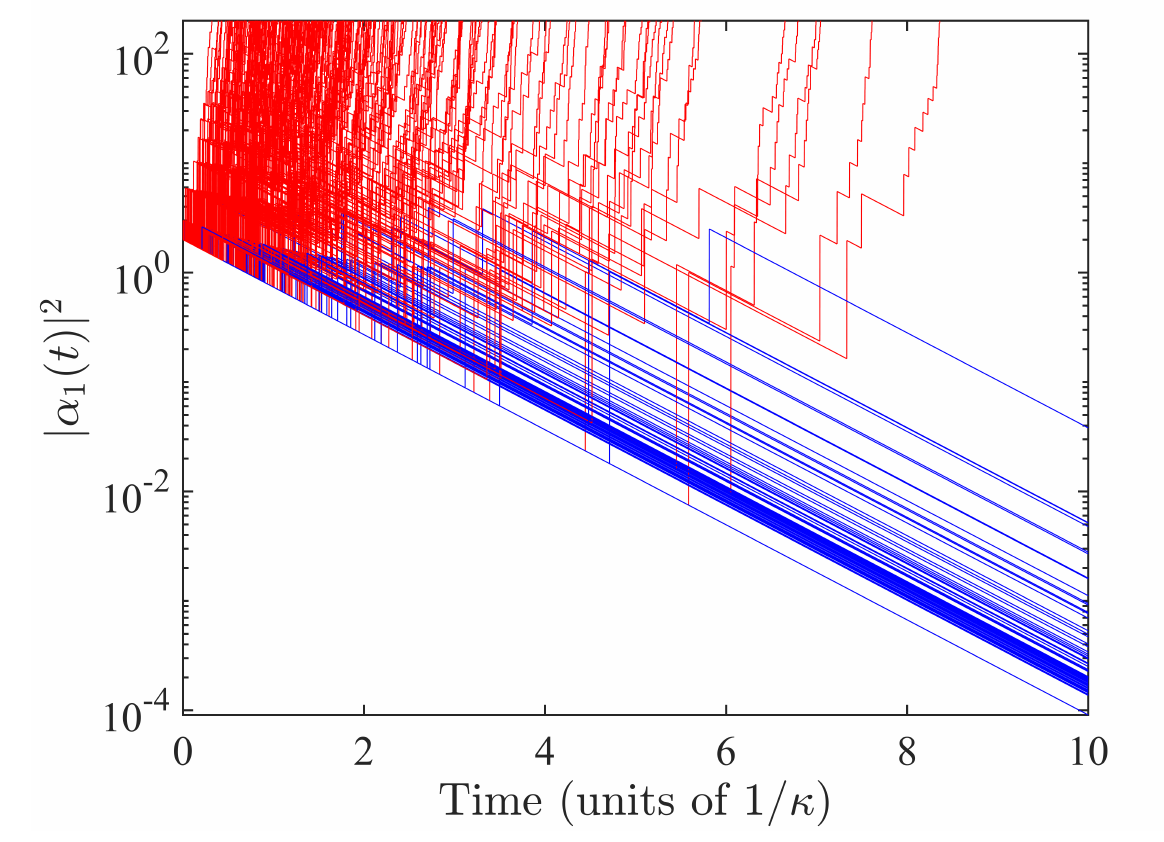}	
	\caption{Illustration of the dynamics of the state $\alpha_2(t)$ in the detector mode basis. Here we present 500 individual trajectories for initial parameters $\gamma_1 (0)=\gamma_2(0)=1$, quantum feedback as described by Eq.~(\ref{QF}) with $\beta_2^{(1)}=1$ and $\beta_1^{(2)}=2$, $\kappa_1 = \kappa_2 = \kappa$, $\Delta t=10^{-3} \kappa^{-1}$ and $\varphi=0$.  We show the population $|\alpha_2(t)|^2$ of the detector mode $a_2$.  It is clear to see the difference in behaviour, as some trajectories diverge, while others decay towards the vacuum, as signified by the different line shades.}
	\label{fig:traj_heatmap}
\end{figure}

In this section, we study the behaviour of the experimental setup in Fig.~\ref{fig:setup} in more detail to understand better how it can be used to estimate the phase $\varphi = \varphi_1 - \varphi_2$. We then consider the fundamental limits of the estimation accuracy that we can expect for the proposed cavity network based on the photon statistics.

\subsection{Dynamics and Quantum trajectories}

In the proceeding discussions we analyse the behaviour and the sensing capabilities of the cavity network shown in Fig.~\ref{fig:setup} for a specific example of quantum feedback, which can be described by
\begin{align} \label{QF}
\underline{\beta}^{(1)} & = \left( \begin{array}{c} 0 \\ \beta_2^{(1)} \end{array} \right) \quad \underline{\beta}^{(2)} = \left( \begin{array}{c} \beta_1^{(2)} \\ 0 \end{array} \right) \, .
\end{align}
As pointed out already in the previous section, we treat the quantum feedback as approximately instantaneous. For simplicity, we also consider the case of perfect photon detection and assume that all photons are counted and trigger feedback pulses. Losses could be incorporated but do not largely effect the overall behaviour of the cavity network and are therefore neglected here \cite{lewisstokes}.  As we shall see below, the performance of the measurement scheme that we propose here does not depend strongly on the exact number of emitted photons. It is therefore also widely independent of the detector efficiency $\eta$, as longs as $\eta$ differs sufficiently from zero, which allows us to only study the case $\eta = 1$ for simplicity.

Although the experimental setup that we analyse in this paper remains always in a coherent state, its dynamics are nevertheless non-trivial.  While optical cavities with continuous laser driving smoothly evolve into a steady state, the same does not always apply in the presence of quantum feedback.  For example, when the feedback is in the form of strong laser pulses, the free decay of the cavity field is perturbed by `kicks' to the dynamics. These kicks occur more often when there are more photons inside the resonator and hence result into a divergence of the average photon number. This highly non-linear behaviour prevents us from obtaining a straightforward closed analytic solution to the master equation and its statistical moments, despite the cavities always being in a coherent state.

Instead of studying the ensemble behaviour though, considering the individual quantum trajectories of the system in Fig.~\ref{fig:traj_heatmap} reveals more subtle behaviour.  In particular, we see the creation of two types of dynamics.  In one case we see a divergence of cavity photon numbers, with each feedback pulse making the state even more likely to emit another photon and thus diverging further due to subsequent feedback pulses.  However, we also see trajectories that do not follow this evolution and decay towards the vacuum state, with only a small number of photon emissions.  Generally, after a reasonable amount of time has passed, trajectories do not swap trajectory class and clearly belong to one of two subensembles, which leads to effective ergodicity breaking \cite{lewismaybee}.  This can be understood due to the photon population of the cavity being either extremely large, resulting in a large number of emissions and hence feedback pulses keeping the photon number high, or a very low photon population, making emissions resulting in stimulation of the cavity field very unlikely.
   
\subsection{Quantum Jump Metrology} \label{Sec:QJM}

The amount of information that can be gained from a measurement is quantified by the Fisher information. Let ${\bf x}$ be a string of data of length $N$ with elements $x_i \in {\mathbb{Z}^+}$, while $P_\varphi({\bf x})$ is the probability for this string to occur given a certain $\varphi$.  The Fisher information $F(P_\varphi) $ for such data is defined as
\begin{eqnarray} \label{Eq:Fish_def}
	F(P_\varphi) & \equiv & \sum\limits_{\bf x} P_\varphi({\bf x}) \left[ \partial_\varphi \ln \left( P_\varphi({\bf x}) \right) \right]^2 \nonumber \\
	& = & \sum\limits_{\bf x} \frac{\left[\partial_\varphi P_\varphi ({\bf x}) \right]^2}{P_\varphi ({\bf x})} \, ,
\end{eqnarray}
where we sum over all possible combinations of output data ${\bf x}$. Since $\varphi$ is the unknown parameter to be probed, the probability distribution $P_\varphi({\bf x})$ must be a function of this variable. Now let $\hat{\varphi}$ be an estimator of the unknown parameter $\varphi$. In the case of the cavity network we consider in this paper, this data corresponds to the photon statistics observed over a period of time. The Cram\'er-Rao bound tells us that the minimum uncertainty achievable by such an estimator is bounded by the Fisher information as
\begin{eqnarray}
	\left(\Delta \hat{\varphi} \right)^2 & \ge & \frac{1}{F(P_\varphi)} \, .
\end{eqnarray}
If the $N$ data points are uncorrelated, each contributes an independent amount of information and $\left(\Delta \hat{\varphi}\right)^2$ scales as in Eq.~(\ref{Eq:SQL}).  If however the data possesses correlations, then the information contribution from each value may be beyond linear with respect to the number of data points.  In particular, the correlations in quantum systems can lead to more precise measurements when compared with a classical system with the same number of particles \cite{seth,lloyd,Maccone}.  This can lead to scaling of the form of the Heisenberg limit in Eq.~(\ref{Eq:HL}).

Here we focus on how quantum jumps may induce strong temporal correlations into the dynamics of the quantum trajectories of a single quantum system. These correlations can then be used to realise measurements with outcomes that manifest themselves in the properties of the trajectories rather than through ensemble averages. To see that such measurements are capable of enhanced performance, as pointed out previously in Refs.~\cite{lewisbeige,lewisstokes}, consider Eq.~(\ref{18}), which describes the time evolution of our system. Defining the Lindblad operators
\begin{align}
	L_d = & \sqrt{\kappa_d} \,D^{(a)}_2 (\beta^{(d)}_2) D^{(a)}_1 (\beta^{(d)}_1) a_d
\end{align}
for ease of writing, we can write the master equation as
\begin{eqnarray} \label{Eq:General-master}
	\dot{\rho} & = & \sum\limits_{d=1,2} \left(L_d \rho L_d^\dagger - \frac{1}{2} \left[L_d^\dagger L_d , \rho\right]_+\right) \, .
\end{eqnarray}
When considering a specific quantum trajectory, we unravel the dynamics and subsequently obtain the stochastic master equation
\begin{eqnarray} \label{Eq:Stochasic-master}
	{\rm d} \rho & = & \sum\limits_{d=1,2} \Bigg[ \left( - \frac{1}{2} \left[L_d^\dagger L_d , \rho\right]_+ + \Tr\left(L_d^\dagger L_d \rho\right) \rho \right) {\rm d} t \nonumber \\
	&& + \left(\frac{L_d \rho L_d^\dagger}{\Tr\left(L_d^\dagger L_d \rho\right)} - \rho \right) {\rm d}N_t^{(d)} \Bigg] \, .
\end{eqnarray}
The Poisson increment ${\rm d}N_t^{(d)}$ in this equation is zero for no-emission at time $t$ and one for an emission into the detector mode $d$ \cite{Jacobs_book}.  From Eq.~(\ref{Eq:Stochasic-master}, it is clear that a quantum jump provides a non-linearity in the evolution of a quantum trajectory and thus correlates the subsequent photon statistics.

To see these correlations, consider the Kraus decomposition of the dynamics, with Kraus operators
\begin{eqnarray}
	K_0 = U_{\rm cond} (\Delta t,0) &\approx & \mathbb{1} - \sum\limits_{d=1,2} \left(\frac{1}{2} L_d^\dagger L_d \right) \Delta t \, , \nonumber \\
	K_1 & \approx & \sqrt{\Delta t} \, L_1 \, , \nonumber \\
	K_2 & \approx & \sqrt{\Delta t} \, L_2 \, .
\end{eqnarray}
The coarse-grained time evolution of the system can thus be modelled by applying Kraus operator $K_x$ after each observation $x=\{0,1,2\}$ in the bath every time step $\Delta t$.  The probability of observing a specific sequence of measurement outcomes is thus
\begin{eqnarray} \label{Eq:Traj_prob}
	p(x_1, \dots , x_N) & = & \Tr \left[ \left(\prod\limits_{i=1}^N K_{x_{N+1-i}} \right) \rho \left(\prod\limits_{i=1}^N K_{x_i}^\dagger \right) \right] \, . \nonumber \\
\end{eqnarray}
Next, notice that the Kraus operators do not commute, due to the effect of the feedback.  Thus the events and measurements are not independent from one another.  This can be seen more clearly by comparing the probabilities
\begin{eqnarray} \label{Markov1}
	p(x_N \vert x_{N-1}) & = & \frac{\Tr\left[K_N K_{N-1} \mathcal{T}_1^{N-2} \left(\rho\right) K_{N-1}^\dagger K_N^\dagger\right]}{\Tr\left[ K_{N-1} \mathcal{T}_1^{N-2}\left(\rho\right) K_{N-1}^\dagger \right] } \nonumber \\
\end{eqnarray}
and
\begin{eqnarray} \label{Markov2}
	&& \hspace*{-0.8cm} p(x_N \vert x_{N-1} x_{N-2})  \nonumber \\
	&=& \frac{\Tr\left[K_N K_{N-1} K_{N-2} \mathcal{T}_1^{N-3}\left(\rho\right) K_{N-2}^\dagger K_{N-1}^\dagger K_N^\dagger\right]}{\Tr\left[ K_{N-1} K_{N-2} \mathcal{T}_1^{N-3}\left(\rho\right) K_{N-2}^\dagger K_{N-1}^\dagger \right]} \, . \nonumber \\
\end{eqnarray}
Here $p(x_N \vert x_{N-1})$ is the probability of measuring $x_N$ after $x_{N-1}$ in the previous time step and $p(x_N \vert x_{N-1} x_{N-2})$ is the probability of measuring $X_N$ after $x_{N-1}$ and $x_{N-2}$, while the superoperator $\mathcal{T}_i^j$ describes a Markovian evolution from time-step $i$ to $j$.  In general, Eqs.~(\ref{Markov1}) and (\ref{Markov2}) differ, meaning the measurement statistics do not form a Markov chain and thus possess non-trivial correlations \cite{lewisbeige,lewisstokes}.  If the Kraus operators are dependent on the unknown parameter, these correlations may lead to a Fisher information growing faster than linearly and thus potentially resulting in an enhanced sensing precision. As we shall see below, in the case of the cavity network with quantum feedback that we consider here, this is indeed the case.

\section{A simple measurement scheme for estimating phase difference} \label{sec4}

In this section, we finally discuss, how the phase difference $\varphi = \varphi_1 - \varphi_2$ between two pathways of the network shown in Fig.~\ref{fig:setup} can be measured. We propose a simple experimental scheme and analyse its performance by calculating the uncertainty $\Delta \varphi$ based on an ensemble of simulated quantum trajectories.  Throughout this analysis, we assume perfect photon detection, though, as mentioned already earlier, losses are not expected to significantly harm the performance of the scheme \cite{lewisstokes}.

\subsection{The basic protocol}

As Fig.~\ref{fig:traj_heatmap} demonstrates, in the presence of quantum feedback, the cavity network in Fig.~\ref{fig:setup} generates two classes of trajectory. One class of trajectories evolves both cavities rapidly to their respective vacuum states, while the other class quickly results in huge photon number populations in both resonators. Our simulations show that the relative size of the subensemble associated with each class has a relatively strong dependence on the phase difference $\varphi$ that we want to estimate. The probability that the total number of photons being emitted within a certain time interval of length $t$ is above a certain threshold therefore acts as a reliable measurement signal for the type of trajectory being observed. For example, a convergent trajectory will have emitted no or few photons until a given time $t$, whereas an eventually divergent one is likely to have emitted many photons due to the repeated pumping of energy into the system. As we shall see below, it is important to choose the experimental parameters and the photon number threshold carefully in order to ensure that useful information is revealed. For example, if the threshold number of photons is too low, it won't faithfully distinguish between the trajectory classes. Moreover, attention needs to be paid to the strength of the feedback and the choice of the initial state. Finding a balance between these parameters is essential when determining the phase difference $\varphi$ as  precisely as possible.

An analytic calculation of the expected measurement signal is not straightforward due to the non-linear dynamics of the cavity network and its lack of a stationary state \cite{lewismaybee}.  Hence, to determine the uncertainty of the above-introduced estimator of $\varphi$ for the chosen measurement signal, we numerically simulate and sample a large number of trajectories over a coarse-grained timescale. This then allows us to estimate the probability of the system emitting a number of photons surpassing the threshold we set, to estimate $\varphi$ as a function of the time $t$. Here we are especially interested in the uncertainty $\Delta \hat{\varphi} (t)$ of this signal.

\begin{figure*}[t]
	\centering
	\includegraphics[width= \textwidth]{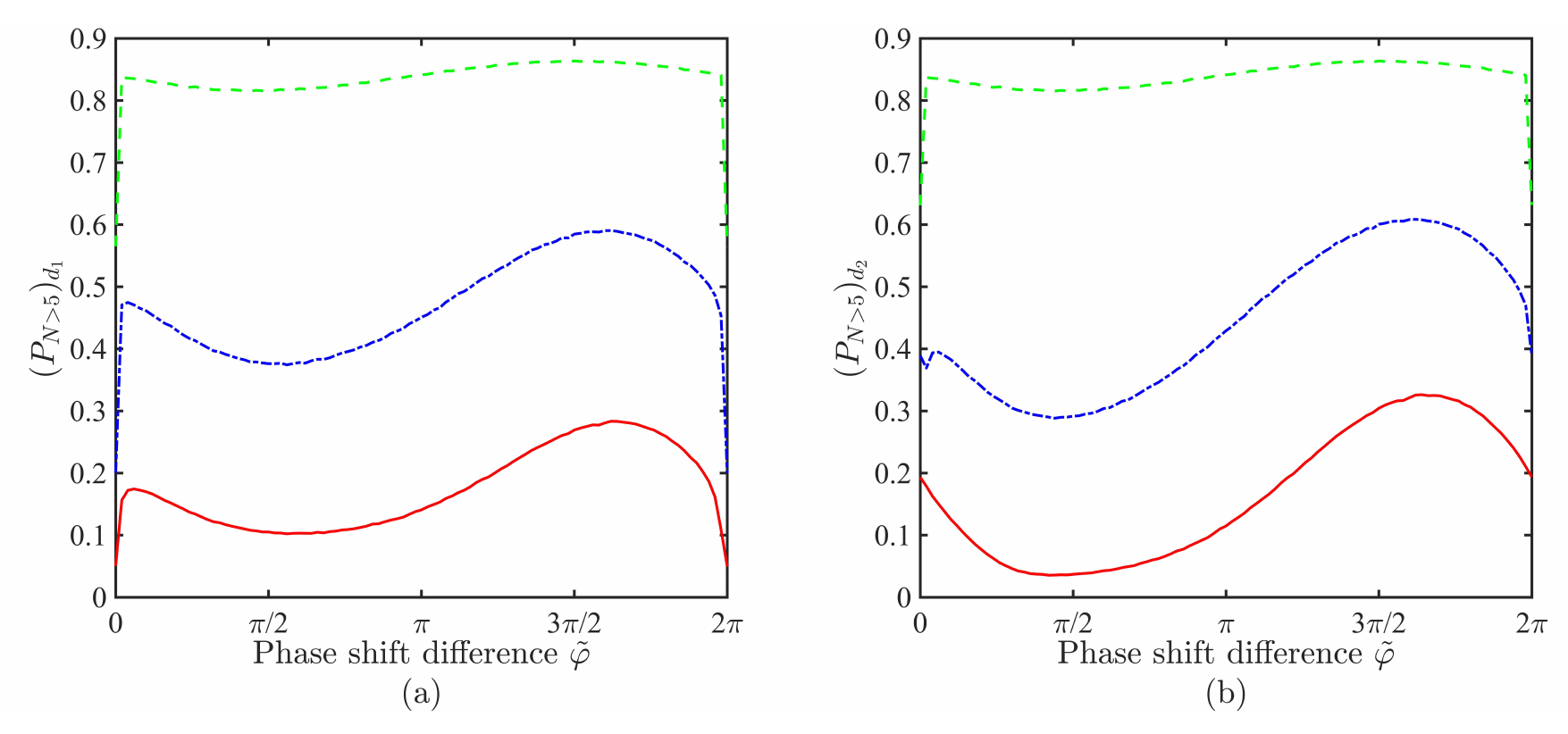}
	\caption{The probability $(P_{N>5})_{d_i}$ with (a) $i=1$ and (b) $i=2$, respectively, of surpassing the threshold number of $N=5$ photon emissions within a certain time interval $(0,t)$ as a function of the phase shift difference $\varphi$. Here $t = 0.5 \, \kappa^{-1}$ (red-solid), $t = \kappa^{-1}$ (blue-dot-dashed) and $t = 10 \, \kappa^{-1}$ (green-dashed). We average over $10^5$ trajectories with $\Delta t=10^{-3} \kappa^{-1}$, $\gamma_1(0) = \gamma_2(0)=1$ and apply quantum feedback as described in Eq.~(\ref{QF}) with $\beta_1^{(2)}=2$ and $\beta_2^{(1)}=1$.}
	\label{fig:estimator}
\end{figure*}

\subsection{Measurement of phase difference}

In the following, we consider a specific, carefully chosen set of parameters to demonstrate the possible quantum advantage of our measurement scheme. As with the threshold value, it is important to choose feedback parameters that are both not too small and not too big. For example, for very weak feedback, we are unlikely to deduce information about $\varphi$ over reasonable timescales. Moreover, for very strong feedback, the dynamics of the cavity network becomes dominated by the feedback, almost all trajectories diverge and the measurement outcomes are essentially independent of the unknown phase that we want to identify. Finally, in order to avoid starting in the vacuum state and for the practicality of implementation, we start each trajectory with a single quantum feedback pulse, which we apply to the vacuum state to prepare a non-trivial initial state.

With all of these factors in mind, Fig.~\ref{fig:estimator} shows three different measurement signals, i.e.~the probability $(P_{N>5})_{d_i}$ of detecting more than $N=5$ photons within a time interval $(0,t)$ at detector $1$ and at detectors 2, respectively, for three different values of $t$. As one would expect, this probability increases as $t$ increases. Moreover, it depends on the phase shift difference $\varphi$, especially when $\varphi$ is close to $0$. Considering Fig.~\ref{fig:estimator}, we see that the optimal phase to conduct an estimation at is around $\varphi = 0$ due to the sharpness of the gradient of the measurement signal at this point.  This corresponds to a crucial point in the dynamics.  When $\varphi = 0$, only one detector mode is ever occupied. However, moving away from this point, the other detector mode begins to be occupied too. Thus, taking advantage of this distinction in the signal allows for the best measurement.  Due to numerical instabilities, evaluating $(P_{N>5})_{d_i}$ exactly at this point however is difficult. We therefore take our data at a nearby value of $\varphi = \pi/10$, where the gradient is still large for reasonable amounts of time.

\subsection{Fisher information for the photon statistics of an optical cavity network}

Before further analysing the performance of the proposed quantum optical sensor, we now calculate the Fisher information of the photon statistics using the methods outlined in Section \ref{Sec:QJM}. In this way, we obtain a bound on the optimum precision that our measurement scheme can achieve. The probability of a certain trajectory with a given number of time steps can be calculated using Eq.~(\ref{Eq:Traj_prob}). To do so, it has to be taken into account that every individual time step has one of four different possible event types, quantified by the probabilities $P_{ij}$ introduced in Section~\ref{sec2}: no-photon, photon only in detector 1, photon only in detector 2 or photons in both detectors. A drawback of this however is that an exact calculation of the Fisher information for a trajectory of length $N$ requires summing over $4^N$ possible trajectories. Hence this approach becomes computationally very challenging for large $N$. Nevertheless, for small $N$, we find scaling beyond linear of the form
\begin{eqnarray} \label{40}
F(N) &\propto & {\cal O}(N^2) - {\cal O}(N) \, .
\end{eqnarray}
This result is in agreement with the scaling behaviour that we previously observed in Ref.~\cite{lewisbeige}, thus demonstrating the presence of correlations in the photon statistics that depend on $\varphi$. 

\begin{figure*}[t]
	\centering
	\includegraphics[width= \textwidth]{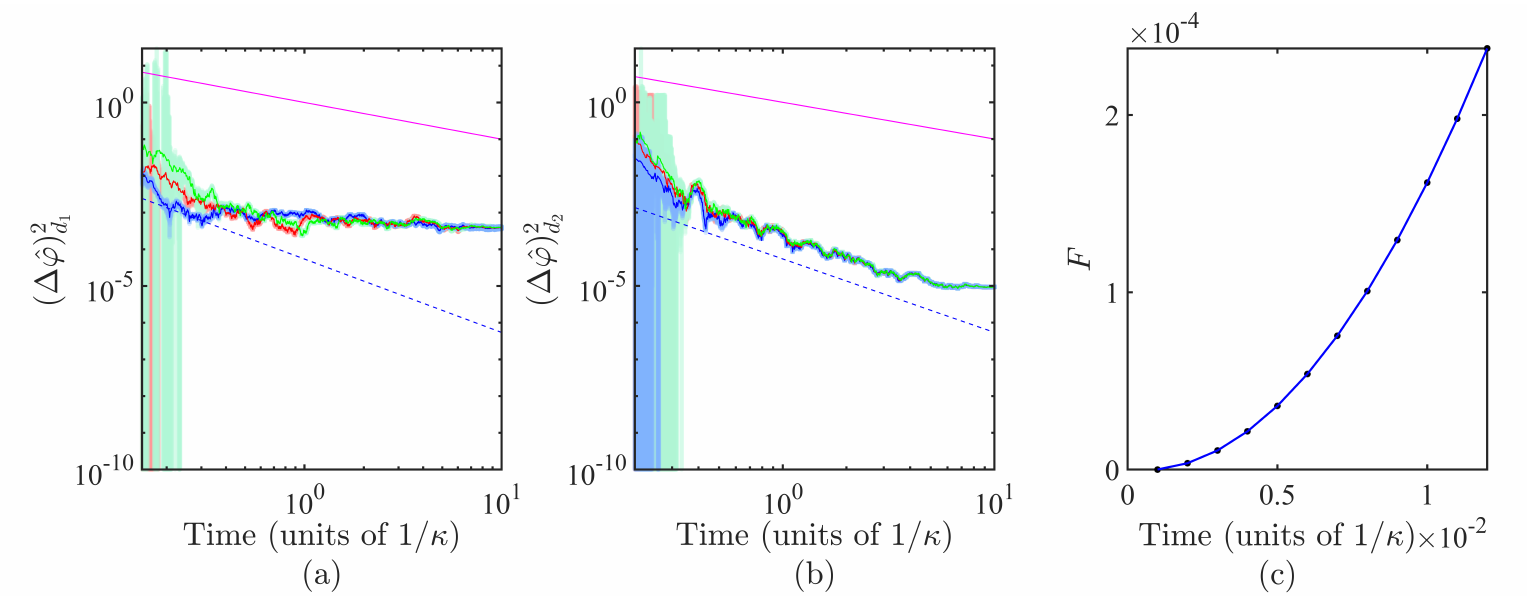}
	\caption{The uncertainty in the phase shift estimation for phase differences $\varphi =\pi/10$ as a function of time for detectors 1 and 2 (in (a) and (b) respectively), for three different thresholds: $N=3$ (blue-lower-solid), $N=5$ (red-upper-solid) and $N=7$ (green-middle-solid).  The probabilities shown in Fig.~\ref{fig:estimator} are generated with $10^4$ trajectories. The error in the simulation, that is determined by evaluating the variance in the uncertainty of phase estimation in 10 subensembles, is shown by the shaded area around each curve. The black dashed line shows the extrapolated value of the reciprocal of the Fisher information, with exact points. The subsequent fit shown in (c).  The Fisher information is obtained exactly for 12 time steps $\Delta t = 10^{-3} \kappa^{-1}$ and the subsequent fit is extended for all time.  This provides an estimated lower bound on the potential sensitivity of the data.  Meanwhile, scaling according to the SQL (magenta-solid) is shown for illustrative purposes.  All system parameters are as in the previous plots.}
	\label{fig:scaling1}
\end{figure*}

The limitation of only having exact results for the Fisher information for a small number of time steps means we do not have a strict bound for the system for larger times.  Instead, we extrapolate the scaling shown for short times.  However we nevertheless expect that this approach provides an upper bound on the Fisher information, as it is likely that at large times the scaling will reduce rather than increase due to a breakdown in correlations between far away time steps. This observation suggests that the estimated bound can nevertheless be useful when comparing it to the uncertainty of our proposed measurement scheme.  We note that general methods of obtaining more accurate estimations of the Fisher information in Eq.~(\ref{40}) would require being able to obtain solutions to the master equation of the cavity network and to determine the stationary state of its dynamics \cite{Gammelmark}, neither of which are present within our system.

\subsection{Sensing performance}

In addition to calculating $(P_{N>3})_{d_i}$ with $i=1,2$ by averaging over many trajectories, we determine the estimation uncertainty $\Delta \hat{\varphi}$ in the following with the help of the standard error propagation formula
\begin{eqnarray} \label{Eq:error}
	\left(\Delta \hat{\varphi} \right)^2 &=& \frac{  \left(\Delta O\right)^2 }{\left| \frac{\partial  \left\langle  O \right\rangle  }{\partial \varphi} \right|^2} \, ,
\end{eqnarray}
where $O$ denotes the relevant observable. The variance in the numerator of this equation is obtained by sampling over a number of subensembles of trajectories, while the visibility in the denominator is obtained numerically.  Utilising this technique, we can study how the error in estimating the phase difference behaves for the sensing protocol that we propose in this paper. 

Fig.~\ref{fig:scaling1} shows the uncertainty $(\Delta \hat{\varphi})^2$ for both detector modes with different threshold photon numbers. In both detector modes, we see that the exact value of the threshold only has a small effect on the obtained results.  This applies, as given even a small number of emissions, the total number of photons in the system is likely to diverge.  We also plot the extrapolated bound obtained from the Fisher information for comparison.  We see that at early times, we get close to the bound in the uncertainty $\Delta \hat{\varphi}$ for detector 1, which then plateaus at later times. For the specific values chosen here, the initial performance of Detector 2 is worse, but $\Delta \hat{\varphi}$ of Detector 2 then steadily decreases and eventually beats that of Detector 1, before plateauing much later.  For both detectors and all three thresholds, we get close to the projected bound with our estimation strategy, thus suggesting that the proposed bound is feasible and our estimation strategy is effective. 

Furthermore, an error analysis for the simulation is carried out by evaluating the variance of the uncertainty in phase shift estimation 
across multiple subensembles of individual trajectories. Fig.~ \ref{fig:scaling1} displays the error in the stochastic simulation, which is represented as a shaded area surrounding the curves. It starts off large due to the strong effect of fluctuations in the early dynamics, before decreasing through time.  However, they progressively decrease as the effect of the quantum feedback on the system becomes more prominent and as more photons are detected in one of the output detectors.

\begin{figure*}[t]
	\centering
	\includegraphics[width=1\textwidth]{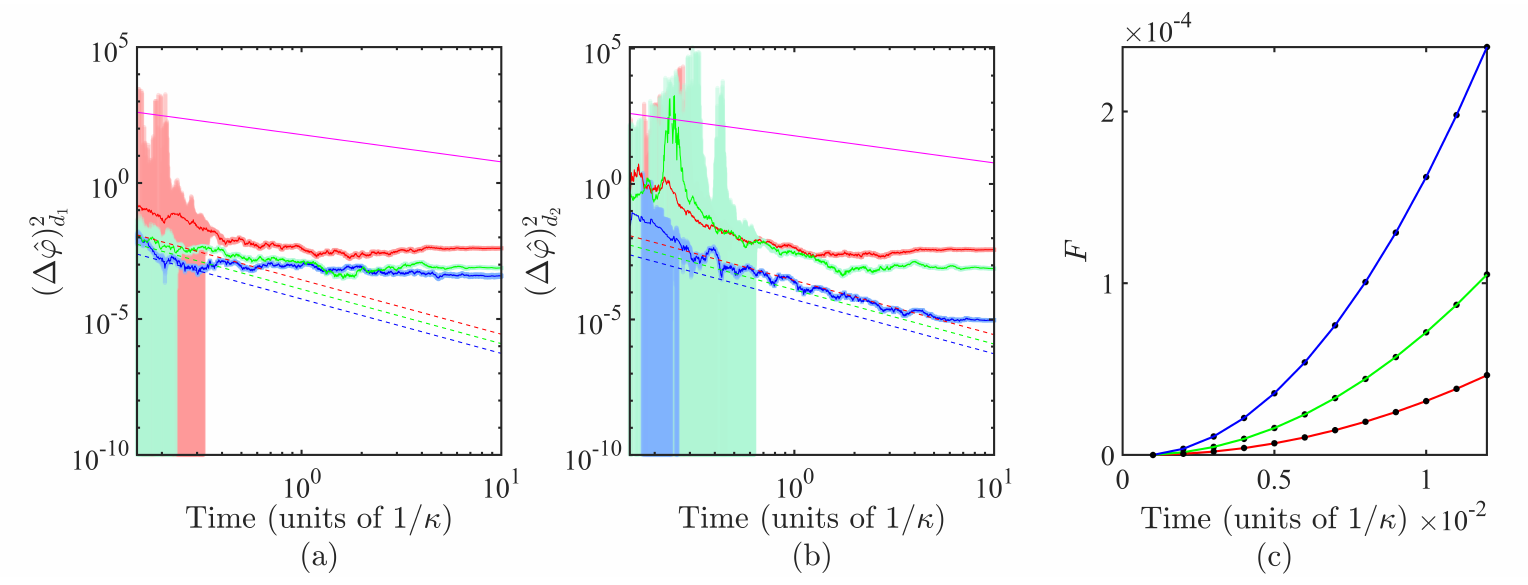}
	\caption{The uncertainty in phase estimation $\Delta \hat{\varphi}$ for detectors 1 and 2 in (a) and (b) respectively, as a function of time. In contrast to Fig.~\ref{fig:scaling1}, the threshold number is fixed to $N=3$ and we now vary the phase $\varphi$ that is estimated. Specifically, we consider $\varphi =\pi/10$ (blue-lower-solid), $\varphi =\pi/10+\pi/2$ (red-upper-solid) and $\varphi =\pi/10+3\pi/4$ (green-middle-solid). All other parameters and generation data is the same as before. The phases closer to $\varphi = 0$ have the best performance due to the gradient generally being steepest in this region. The Fisher information obtained exactly is shown again in (c) for each phase, with the extrapolation of it shown in the bounds in (a) and (b) as the dashed lines.  We again show the SQL-scaling for illustrative purposes in the magenta line.}
	\label{fig:scaling2}
\end{figure*} 

Next we have a closer look at the performance of the proposed sensing scheme for different values of $\varphi$. Fig.~\ref{fig:scaling2} shows $\Delta \hat{\varphi}$ for both detectors as a function of time. As expected, we find that the sensor performs best when $\varphi$ is close to $0$. This result is confirmed when looking at the bounds for different values of $\varphi$, which also suggest that the distinguishability of the photon statistics is sharpest around $\varphi = 0$, although the differences for the different phases are relatively small. For completeness, let us also point out that we have seen already in Fig.~\ref{fig:estimator}, the gradient of the signal is generally the sharpest at this point.

We find that the sensor performance can approach the projected bound for large parts of the evolution when evaluating at $\varphi = \pi/10$ with a threshold of $N=3$, as seen in Fig.~\ref{fig:scaling2}, for the parameters chosen here. This is a promising result, as it justifies the projection of the bound by following the same trend, even if the exact values do not match.  It is also likely that the signal we use as an estimator here is not optimal and as such we would not expect it to fully saturate the bound.  Moreover, the parameter choices are not necessarily optimised, meaning other regimes may yield stronger results.  Nevertheless, the scaling of our signal is promising and therefore is a strong result for demonstrating the quantum enhancement of the quantum jump metrology scheme which we analyse here.

\section{Conclusions} \label{sec5}

In this paper we have demonstrated how the phase difference between the ``arms'' of an optical cavity network (c.f.~Fig.~\ref{fig:setup}) can be inferred, requiring only single photon detectors and quantum feedback in the form of strong but approximately instantaneous laser pulses. Despite the simplicity of the proposed measurement scheme, we have shown that it is capable of estimating the phase difference with a sensitivity beyond the standard quantum limit. As in previous work \cite{lewisbeige,lewisstokes}, the presence of quantum jumps in the form of photon emissions is continuously monitored. Subsequent feedback on the cavities creates non-linearities in the system dynamics, thereby inducing correlations in the photon statistics observed by the detectors. 

In the experimental setup that we consider here, the cavities remain always in a coherent state. Although photon emission does not alter the state of the cavities directly, it reveals information about the state of the resonators. Similarly, not observing photons reveals information. The dependence of the quantum feedback induced dynamics on the parameter that we want to measure leads to effective ergodicity-breaking in the dynamics of the system, similar to the ergodicity breaking discussed in Ref.~\cite{lewismaybee} and results in two different classes of trajectory. In this paper, we have shown that measuring the probability for these two classes of dynamics to occur reveals information about the phase difference between the arms of the cavity network beyond the standard quantum limit. Even better scaling might be achieved by preparing the cavities in a more complex initial state than a coherent state. However, the use of coherent states offers experimental simplicity and as such the proposed scheme can be operated more straightforwardly and for longer periods of time.

A limitation of our work is that the results are only attainable numerically, and the Fisher information is only calculable exactly for small times due to its exponential growth in the number of probes. However, the expected behaviour has been predicted with approximations previously and the observed behaviour here is in line with previous work \cite{lewisbeige}. Moreover, it may be possible to obtain estimates for the Fisher information at large times by using sampling techniques.  Even in this case however, the number of possible trajectories is extremely large even for reasonable values of $N$, thus meaning the number of trajectories needed to sample over may also need to be large.  Due to how well the projected bound matches the uncertainty predicted from our measurement signal though, we believe our fitting of the Fisher information to be reasonable.

An alternative measurement protocol for gaining information about the phase would be to utilise all information gained in the continuous monitoring of the cavity network and follow a Bayesian inference procedure.  In an open system where the photon statistics are observed this is perhaps the most natural way to envisage the measurement of an unknown parameter \cite{Kay,Gammelmark2,Opto}. This could be further supplemented by a strong quantum measurement of the cavity state at the end of the observation which supplements the information gained from monitoring the photon statistics \cite{Albarelli}.  For the purpose of this work though, we choose to just consider the Fisher information as a proof-of-principle that a quantum enhancement exists and concentrate on a simple to implement measurement. Because of this property, cavity networks like the one shown in Fig.~\ref{fig:setup} are very likely to attract more attention for applications in quantum sensing and might play a crucial role in the development of quantum machine learning devices.

\subsection*{Acknowledgements}
LAC acknowledges support from the Foundation for Polish Science within the ``Quantum Optical Technologies" project carried out within the International Research Agendas programme co-financed by the European Union under the European Regional Development Fund.  KAR acknowledges the support from the Ministry of Higher Education, Research and Innovation in the Sultanate of Oman funded by The National Postgraduate Scholarship Programme.

\end{document}